\documentclass[12pt]{iopart}
\usepackage{bm}
\usepackage{amsmath}
\usepackage{epsfig}
\usepackage{rotating}
\usepackage{graphicx}
\usepackage{url}

\def\be{\begin{equation}}
\def\lan{\left\langle}
\def\ran{\right\rangle}
\def\ee{\end{equation}}
\def\barr{\begin{array}}
\def\earr{\end{array}}

\def\l{\left}
\def\r{\right}
\def\dis{\displaystyle}
\def\f{\frac}
\def\cc{{\cal C}}

\def\ca{{\cal A}}

\def\spin{\frac{1}{2}}

\def\cf{{\cal F}}
\def\la{\lambda}
\begin{document}

\title{Proxy-$SU(4)$ symmetry in A=60-90 region}

\author{V.K.B. Kota$^1$, R. Sahu$^2$}
\address{$^1$Physical Research  Laboratory, Ahmedabad 380 009, India}
\ead{vkbkota@prl.res.in}
\address{$^2$National Institute of Science and Technology, Palur Hills, Berhampur-761008, India}
\vspace{10pt}
\begin{indented}
\item[]November 2023
\end{indented}

\begin{abstract}

Applications of the  proxy-$SU(3)$ model of Bonatsos and collaborators to nuclei in 
A=60-90 region introduces proxy-$SU(4)$ symmetry. Shell model spaces with 
single particle (sp) orbits $^1p_{3/2}$, $^1p_{1/2}$, $^0f_{5/2}$ and $^0g_{9/2}$
are essential for these nuclei and also protons and neutrons in this region
occupy the same sp orbits. With this and applying the "proxy scheme", the $^0g_{9/2}$ changes
to $^0f_{7/2}$ giving the SGA $U(40) \supset [U(10) \supset G \supset 
SO(3)] \otimes [SU(4) \supset SU_S(2) \otimes SU_T(2)]$. With $G=SU(3)$, we have the
proxy-$SU(3)$ model. It is easy to
see that proxy-$SU(3)$ symmetry implies goodness of the $SU(4)$ symmetry appearing above,
i.e. proxy-$SU(4)$ symmetry. Shell model calculations pointing out the need for
$^0g_{9/2}$ orbit, ground state masses, shape changes and shape co-existence in
A=60-90 region and GT distributions clearly show the importance of proxy-$SU(4)$ in this
mass region. Besides presenting this evidence, new proxy schemes with $G=SU(5)$,
$SO(6)$ and $SO(10)$ that are generated by good proxy-$SU(4)$ 
symmetry are described in some detail. An important feature is that the four proxy symmetries $SU(3)$, $SO(6)$, $SU(5)$ and $SO(10)$ appear twice. 

\end{abstract}

\vspace{10pt}
\noindent{\it Keywords}: Proxy-SU(3), Proxy-SU(4), A=60-90 region, Nuclear structure

%

\section{Introduction}
Introduction of isospin in 1932 by Heisenberg \cite{isospin}, spin-isospin $SU(4)$ algebra in 1937 by 
Wigner \cite{Wig-37}, seniority quantum number by Racah with pairing $SU(2)$ algebra in 1943 \cite{Racah} and
the rotational $SU(3)$ algebra in 1958 by Elliott \cite{Ell-58a,Ell-58b}
are the four most significant
developments in the subject of 'symmetries in nuclei'. This area of research in nuclear structure physics has its first renaissance with the development of the Interacting Boson Model (IBM) by Arima and Iachello \cite{Iac-1,Iac-2} in late 70's and 80's. More importantly, the subject of 'symmetries in nuclei' is undergoing a second renaissance in the last 5-10 years with several new directions in this subject are being explored. Some of these are (i) point group symmetries in heavy nuclei such as $^{152}$Sm, $^{156}$Gd, $^{236}$U \cite{Dudek-1,Dudek-2,Piet} and in lighter nuclei such as $^{12}$C, $^{13}$C and $^{16}$O \cite{Bij-1,Bij-2,Bij-3}, (ii) symmetries for shape coexistence \cite{Leviatan}, (iii) multi-configuration or composite symmetries for cluster states \cite{Cseh}, (iv) symmetry adopted no-core shell model based on $SU(3)$ and $Sp(6,R)$ algebras \cite{JPD-1,JPD-2,JPD-3,JPD-4}, (v) proton-neutron $Sp(12,R)$ model \cite{Ganev-1,Ganev-2,Ganev-3}, (vi) multiple algebras in shell model and IBM spaces giving for example multiple pairing and $SU(3)$ algebras \cite{Kota-1,Kota-2}, (vii) pairing algebras with higher
order interactions, generalized seniority and seniority isomers \cite{Pfeng,Mahe-1,Mahe-2,Kota-2}, (viii) proxy-$SU(3)$ scheme within shell model \cite{Bonat-1,Bonat-rev},
(ix) symmetry restoration methods in mean-field theories \cite{Ring}, (x) new techniques for obtaining Wigner coefficients involving $SU(3) \supset SO(3)$, $SO(5) \supset SO(3)$ and $SU(4) \supset SU(2) \otimes SU(2)$ \cite{Feng1,Feng2,suwig} and (xi) studies involving Bohr Hamiltonian with sextic and other types of potentials for shape phase transitions and shape coexistence pointing out the need for three and higher-body terms in IBM and perhaps also in shell model \cite{Bud-1,Bud-2,Bud-3,Bud-4,Bud-5,Bud-6}. (xii) random matrix ensembles with Lie algebraic symmetries for quantum chaos in nuclei \cite{MK-10,EE-book,EE-pnege}. In this article we consider proxy-$SU(4)$ symmetry that follows from the proxy-SU(3) model when applied to  nuclei with valence nucleons in 28-50 shell. Details of the proxy-$SU(3)$ model are reviewed in \cite{Bonat-rev}.

Applications of the  proxy-$SU(3)$ model to nuclei in 
A=60-90 region introduces proxy-$SU(4)$ symmetry. Shell model spaces with single particle (sp) orbits $^1p_{3/2}$, $^1p_{1/2}$, $^0f_{5/2}$ and $^0g_{9/2}$ are essential for these nuclei and also protons and neutrons in these region occupy the same sp orbits. With this and applying the "proxy scheme", the $^0g_{9/2}$ changes to $^0f_{7/2}$ giving the SGA $U(40) \supset [U(10) \supset G \supset SO(3)] \otimes [SU_{ST}(4) \supset SU_S(2) \otimes SU_T(2)]$. With $G=SU(3)$, we have proxy-$SU(3)$ model. It is easy to see that the proxy-$SU(3)$ implies goodness of the $SU(4)$ symmetry appearing above, i.e. proxy-$SU(4)$ symmetry.
Besides this, many other properties of A=60-90 nuclei give indirect evidence for the application of proxy-$SU(4)$ symmetry
in this region. These include for example, ground state masses, shape co-existence and prolate-oblate transitions in A=60-90 region, role of $^0g_{9/2}$ orbit in spectroscopy such as triple-forking in $^{68}$Ge, Gamow-Teller distributions and so on. With proxy-$SU(4)$, it is possible to have not only proxy-$SU(3)$ symmetry but also other proxy symmetries with $G=SO(6)$, $SU(5)$ and $SO(10)$. Let us add that in the $^1p_{3/2}$, $^1p_{1/2}$, $^0f_{5/2}$ and $^0g_{9/2}$ space, it is possible to use pseudo-$SU(3)$ with pseudo-$SU_{ST}(4)$ in $^1p_{3/2}$, $^1p_{1/2}$, $^0f_{5/2}$ space plus proton-neutron pairing $SO(5)$ algebra in $^0g_{9/2}$ space. Such an analysis may also give
good results \cite{psu3,psu4}. Our interest in this paper is in exploring proxy- $[U(10) \supset G \supset SO(3)] \otimes SU_{ST}(4)$ symmetries and their applications. Now we will give a preview.

In Sections 2, proxy-$SU(3)$ model is briefly introduced. Also its application to A=60-90 nuclei, that necessarily requires good proxy-SU(4) symmetry, is discussed. In Section 3 of the paper,  further evidence for proxy-$SU(4)$ will be discussed using several examples in A=60-90 region. In Sections 4 and 5 we will describe the new $SO(6)$, $SU(5)$ and $SO(10)$ proxy schemes in A=60-90 region that arise due to good proxy-$SU(4)$ symmetry. In Section 4 the group generators and quadratic Casimir invariants are given. Similarly, in Section 5 results for reductions of various group irreducible representation (irreps) to their subgroup irreps are discussed. These results will 
allow one to identify the ground state structure and typical spectra generated by the three new symmetry limits. Finally, Section 6 gives conclusions and future outlook.

\section{Proxy-$SU(3)$ model application in A=60-90 region and the need for proxy-$SU(4)$ symmetry}

\subsection{Proxy-$SU(3)$ model}

Bonatsos, Casten and others observed \cite{Bonat-1,Bonat-rev} that enhanced proton-neutron
interactions occur in heavy deformed nuclei with protons and neutrons occupying
different shells and with the valence number of protons is approximately same as
the valence number of neutrons, i.e. Z$_{val} \sim$ N$_{val}$. 
More
strikingly, it is observed that in these nuclei the valence protons and neutrons
fill synchronously the Nilsson orbits $\l[\eta n_z \Lambda\r]\Omega$ such that
$\l[\Delta{\eta}, \Delta{n_z}, \Delta{\Lambda}\r]\Delta{\Omega} = [110]0$. For example, for Z=50-82
proton shell and N=82-126 neutron shell, one has the successive $p-n$ orbit combinations:
$[431]1/2-[541]1/2$, $[422]3/2-[532]3/2$, $[413]5/2-[523]5/2$,
$[420]1/2-[530]1/2$ and so on. With this correlation, one can replace the intruder high-$(\ell, j)$ orbit that is pushed down due to strong spin-orbit force by the proxy $(\ell-1,j-1)$ orbit
by ignoring the high-lying $(\ell,j:k=\pm j)$ states. Then for example $^0h_{11/2}$ in 50-82 shell changes to proxy $^0g_{9/2}$ and then 50-82 shell becomes proxy $\eta=4$ shell. Similarly,
$^0i_{13/2}$ in 82-126 shell changes to $^0h_{11/2}$ and then 82-126 shell becomes proxy $\eta=5$. With the proxy oscillator shells, we have $SU(3)$ symmetry in each shell and by coupling the proxy-$SU(3)$ algebras in the valence shells for the valence protons and neutrons in heavy nuclei, we have proxy-$SU(3)$ symmetry. In the proxy model
for example, for $^{168}$Er we have 18 protons in 50-82 shell with proxy
$\eta=4$ shell giving the leading proxy $SU(3)$ irrep to be $(\lambda_p \mu_p)=(18,6)$. Similarly, it has 18 neutrons in 82-126 shell with
proxy $\eta=5$ shell giving the leading proxy $SU(3)$ irrep to be $(\la_n
\mu_n)=(36,6)$; see \cite{Bonat-rev} for details. Then, the leading $SU(3)$ irrep $(\la_p +
\la_n, \mu_p + \mu_n)$ for the total system is $(54,12)$. On the
other hand, in the pseudo-SU(3) model for heavy nuclei we have two pseudo oscillator shells each with a
$SU(3)$ algebra and in addition the two intruder single-$j$ orbits that are treated
differently \cite{JPD-Weeks}. Therefore, the $SU(3)$ irreps in the pseudo-$SU(3)$ model will be
different from the $SU(3)$ irreps in the proxy-$SU(3)$ model and thereby they generate in some situations different physical properties. For example, proxy-$SU(3)$ model predicts prolate dominance over oblate shapes in heavy nuclei and also the onset of oblate deformation; see Sect. 2.2 and also \cite{Bonat-rev}.
Before proceeding further, it is useful to mention that for rare earth nuclei with valence protons in 50-82 shell give proxy $\eta_p=4$ shell with the SGA 
$$
U_p(30) \supset \l[U_p(15) \supset SU_p(3) \supset SO_{L_p}(3)\r] \otimes SU_{S_p}(2)
$$
and similarly, for neutrons in 82-126 shell give proxy $\eta_n=5$ shell with the SGA
$$
U_n(42) \supset \l[U_n(21) \supset SU_n(3) \supset SO_{L_n}(3)\r] \otimes SU_{S_n}(2)\;.
$$  
Now, one can couple the two $SU(3)$'s giving $SU_{p+n}(3) \supset SO_{p+n}(3)$ and the spins $S_n$ and $S_n$ to give total $S$. This generates the proxy-$SU(3)$ group chain. Similarly, for actinides we have proxy $\eta_p=5$ and $\eta_n=6$ shells. Our interest in the present paper is in applying the proxy-shells and proxy-$SU(3)$ concept to A=60-90 nuclei. Here the active sp orbits are $^1p_{3/2}$, $^1p_{1/2}$, $^0f_{5/2}$ and $^0g_{9/2}$ and more importantly the valence protons and neutrons in this region occupy the same sp orbits. Therefore, the spin-isospin $SU(4)$ algebra/symmetry comes into play. This new aspect will be considered in the
following subsection.   

\subsection{Proxy-$SU(3)$ and $SU(4)$ algebras in A=60-90 region}

With $^{56}$Ni core, for A=60-90 nuclei the active sp orbits are
$^1p_{3/2}$, $^1p_{1/2}$, $^0f_{5/2}$ and $^0g_{9/2}$. We will discuss the strong evidence for the importance of $^0g_{9/2}$
in Section 3. With valence nucleons occupying these four orbits, it is not possible to have $SU(3)$ symmetry. However, if we use the proxy-shells concept, the $\l.\l|^0g_{9/2}, k=\pm 9/2\r.\ran$
states are ignored and assume that the remaining states belong to the proxy $^0f_{7/2}$ orbit, we have the proxy $\eta=3$ shell with proxy $\ell$ values $1$ and $3$. Then, for nuclei with valence protons and neutrons both in the proxy $\eta=3$ shell, the SGA is $U(40)$. For $SU(3)$ algebra in this space we need $U(40) \supset U(10) \otimes SU_{ST}(4)$ and the spatial part $U(10)$, generated by nucleons in proxy $\ell=1$ and $3$ orbits, will admit $SU(3)$ subalgebra giving the group-subgroup chain
\be
SU(3)\;\; limit:\;\; U(40) \supset \l[U(10) \supset SU(3) \supset SO_L(3)\r] \otimes \l[SU_{ST} \supset SU_S(2) \otimes SU_T(2)\r]\;.
\label{eq.prox1}
\ee
Using LST coupling, it is easy to see that $U(40)$ is generated by the operators
\be
u^{L,S,T}_{\mu_\ell , \mu_S , \mu_T}(\ell_1,\ell_2) = \l(a^\dagger_{\ell_1 \spin \spin} \tilde{a}_{\ell_2, \spin, \spin}\r)^{L,S,T}_{\mu_\ell , \mu_S , \mu_T}\;;\;\;\; \ell_1,\ell_2=1,3\;.
\label{eq.proxa}
\ee
Note that
$$
a_{\ell m_\ell\;; \spin m_s\;;\spin m_t} = (-1)^{m_\ell - m_s -m_t}\,\tilde{a}_{\ell-m_\ell\;; \spin -m_s\;;\spin -m_t}\;.
$$
Now, the generators of $U(10)$ and $SU_{ST}(4)$ are
\be
\barr{l}
U(10)\;:\;\; u^{L,0,0}_{\mu_\ell , 0, 0}(\ell_1,\ell_2)\;;\;\;\;\ell_1,\ell_2=1,3\;,\\
SU_{ST}(4)\;:\;\;\;X^{S_0,T_0}_{\mu , \mu^\prime} = \dis\sum_{\ell} \sqrt{2\ell+1}\; u^{0,S_0,T_0}_{0, \mu, \mu^\prime}(\ell,\ell)\;;\;\;\;(S_0T_0)=(10),(01),(11)\;.
\earr \label{eq.proxb}
\ee
By including $X^{0,0}$ the number operator we will have $U_{ST}(4)$ generators. Similarly, the $(S_0T_0)=(10)$, $(01)$ $(11)$ correspond
to spin, isospin and Gamow-Teller (GT) operators. Finally, the generators of $SU(3)$ are the angular momentum ($L$) and quadrupole moment ($Q^2$) operators,
\be
\barr{l}
L^1_q = 2\dis\sqrt{2}\,u^{1,0,0}_q(1,1)\; + \;4\dis\sqrt{7} \, u^{1,0,0}_q(3,3)\;,\\
Q^2_\mu(\alpha) = -\dis\f{18}{5} \dis\sqrt{6}\,u^{2,0,0}_\mu(1,1) -\dis\f{36}{5} \dis\sqrt{\f{7}{3}}\,u^{2,0,0}_\mu(3,3)
+\alpha\,\dis\f{12}{5}\dis\sqrt{14}\,\l\{u^{2,0,0}_\mu(1,3) + u^{2,0,0}_\mu(3,1)\r\}\;;\\
\alpha=+1\;\;\mbox{or}\;\;-1\;.
\earr \label{eq.proxc}
\ee
Thus, there are two proxy-$SU^{\alpha}(3)$ algebras and they correspond to $\alpha=+1$ and $-1$. It is useful to note that the
quadratic Casimir invariant $\cc_2(SU^{(\alpha)}(3))$ of $SU(3)$ is related to $Q^2(\alpha) \cdot Q^2(\alpha)$ operator,
\be
\f{1}{4}\,Q^2(\alpha) \cdot Q^2(\alpha) = \cc_2(SU^{(\alpha)}(3)) - \f{3}{4}\,L.L\;\;,
\label{eq.c2qq} 
\ee
and its eigenvalues are $\lan \cc_2(SU^{\alpha}(3)) \ran^{(\lambda,\mu)} = \l[\lambda^2 + \mu^2 + \lambda \mu + 3(
\lambda + \mu)\r]$. Thus, the $SU(3)$ spectrum, generated by $Q^2(\alpha) \cdot Q^2(\alpha)$ will not depend on $\alpha$. However, the value of $\alpha$ matters while considering transition strengths. For example, if the $\alpha$ used in constructing the $SU(3)$ states and the $\alpha$ used in the quadrupole transition operator for calculating quadrupole transition matrix elements are different (i.e. one is $+1$ and other $-1$ or vice-verse). See \cite{Kota-1} for further discussion on this important aspect of the $SU(3)$ limit. In the following discussion we assume that the $\alpha$ value is same for both the $SU(3)$ states and for the quadrupole transition operator.  

Applying Eq. (\ref{eq.prox1}), it is easy to see that goodness of proxy-$SU(3)$ symmetry in the A=60-90 region
implies good proxy-$SU(4)$ symmetry. In order to apply the scheme
given by Eq. (\ref{eq.prox1}), we need the enumeration of the
quantum numbers for the irrep labels defining the basis states for $m$ nucleons in this space, i.e. the states 
$$
\l.\l|m,\{f\} (\lambda,\mu) KL;\{F\}ST;\vec{J}=\vec{L}+\vec{S}; M_T\r.\ran
$$
where $m$ is nucleon number defining the irrep $\{1^m\}$ of $U(40)$, $\{f\}$ are the irreps of $U(10)$, $(\lambda,\mu)$ are irreps of $SU(3)$, orbital angular momentum quantum number is $L$ (also the $K$ quantum number that
is needed \cite{Ell-58a,su3-book} is shown), $\{F\}$ are the irreps of $U_{ST}(4)$ and they contain spin $S$, isospin $T$ and its $z$-projection $M_T=(m_p-m_n)/2$. Note that $m_p$ and $m_n$ are valence proton and neutron numbers and $m=m_p+m_n$. Firstly, as all the $m$ nucleon states, $\binom{40}{m}$ in number, must be antisymmetric, they all
belong to the irrep $\{1^m\}$ of $U(40)$. Moreover, the antisymmetry requirement implies $\{f\}$ of $U(10)$ must be the transpose of $\{F\}$ of $U(4)$ and then,
\be
\barr{l}
\{F\} = \{F_1,F_2,F_3,F_4\}\;,\;\;\;F_1 \ge F_2 \ge F_3 \ge F_4 \ge 0\;\;\;\mbox{and}\;\;\;m=F_1+F_2+F_3+F_4\;, \\
\Rightarrow\;\;\;\{f\}= \{4^a 3^b 2^c 1^d\}\;\;\;\mbox{with} \\
a=F_4,\;\;b=F_3-F_4,\;\; c=F_2-F_3,\;\; d=F_1-F_2\;.
\earr \label{eq.prox2}
\ee
Note that the $SU(4)$ irrep corresponding to $\{F_1,F_2,F_3,F_4\}$ is $\{F_1-F_4,F_2-F_4,F_3-F_4\}$ (some authors use some other
equivalent notations \cite{JCP,MK-10,Piet-su4}). 
For us with Eq. (\ref{eq.prox1}), the oscillator quantum  number is $\eta=3$ and then $M_T$ values for $m \leq 20$ are
$|M_T|=m/2$, $m/2-1$,\ldots, $0$ or $1/2$ and for $m > 20$, we have $|M_T|=(40-m)/2$, $(40-m)/2-1$,\ldots, $0$ or $1/2$.
Proceeding further, given the nucleon number $m$ and the isospin $T=|M_T|$, the lowest $U(4)$ irrep $ \{F^0_1,F^0_2,F^0_3,F^0_4\} $,
that is of great interest in applications, is given as follows \cite{JCP,MK-10,Piet-su4}. For $m$ even we have
\be
\barr{l}
\{F^0_1,F^0_2,F^0_3,F^0_4\} = \l\{\frac{m+2T}{4}, \frac{m+2T}{4}, \frac{m-2T}{4}, 
\frac{m-2T}{4}\r\}\;\;\;\mbox{for}\;\;\;\frac{m}{2}+T\;\;\mbox{ even}\;,\\
\{F^0_1,F^0_2,F^0_3,F^0_4\} = \l\{\frac{m+2T+2}{4}, \frac{m+2T-2}{4}, 
\frac{m-2T+2}{4}, \frac{m-2T-2}{4}\r\}\;\;\;\mbox{for}\;\;\;
\frac{m}{2}+T\;\;\mbox{odd}\;.\\
\earr \label{eq.prox3}
\ee
The only exception is $T=0$ for $m=4r+2$ type and then 
\be
\{F^0_1,F^0_2,F^0_3,F^0_4\} = \l\{\frac{m+2}{4}, \frac{m+2}{4}, \frac{m-2}{4}, \frac{m-2}{4}\r\}\;.
\label{eq.prox4}
\ee
Similarly, for odd-$m$ we have
\be
\barr{l}
\{F^0_1,F^0_2,F^0_3,F^0_4\} = \l\{\frac{m+2T+2}{4}, \frac{m+2T-2}{4}, \frac{m-2T}{4},
\frac{m-2T}{4}\r\}\;\;\;\mbox{for}\;\;\;\frac{m}{2}+T\;\;\mbox{odd}\;,\\
\{F^0_1,F^0_2,F^0_3,F^0_4\} = \l\{\frac{m+2T}{4}, \frac{m+2T}{4}, \frac{m-2T+2}{4}, 
\frac{m-2T-2}{4}\r\}\;\;\;\mbox{for}\;\;\;\frac{m}{2}+T\;\;\mbox{even}\;.\\
\earr \label{eq.prox5}
\ee
These show that the lowest $SU(4)$ irrep is $\{0\}$ with $T=0$ for N=Z even-even nuclei and it is $\{1,1\}$ with $T=0,1$ for N=Z
odd-odd nuclei. Let us add that the quadratic Casimir invariant of $SU(4)$ is $\cc_2(SU(4)) = S^2 + T^2 + (\sigma\tau) \cdot (\sigma \tau)$. Also, $\cc_2(SU(4)) + \cc_2(U(10))$ is a simple function of the number operator. Thus, the lowest $SU(4)$ irrep implies that the highest $U(10)$ irrep states will be lowest in energy. Then, for N=Z even-even (m=4r), N=Z odd-odd (m=4r+2) and N=Z $\pm$1 (m=4r$\pm$1) odd-A nuclei, the $U(10)$ irreps for the lowest states (with lowest $T$) are $\{4^r\}$, $\{4^4,2\}$, $\{4^r,1\}$ and $\{4^{r-1},3\}$ with $(ST)$ values $(0,0)$, $(01) \oplus (10)$, $(\f{1}{2},\f{1}{2})$ and $(\f{1}{2},\f{1}{2})$ respectively. These lowest $U(10)$ irreps will be used in 
Section 5.

Given the lowest $U(4)$ (or $SU(4)$) irrep for a given $(m,T)$, we have from Eq. (\ref{eq.prox2})
the lowest $U(10)$ irrep $\{f_1,f_2,\ldots,f_{10}\}$. Then the lowest (highest weight) $SU(3)$ irrep $(\lambda_H,\mu_H)$ is given  by the simple formula \cite{Kota-arx},
\be
\lambda_H = \dis\sum_{r=0}^3 \dis\sum_{x=0}^r (3-2r+x) \times 
f_{1+x+\frac{r(r+1)}{2}}\;,\;\;\;\;\mu_H=\dis\sum_{r=0}^3 \dis\sum_{x=0}^r (r-2x) \times f_{1+x+\frac{r(r+1)}{2}}\;. 
\label{eq.prox6}
\ee
The lowest $SU(3)$ irrep for
all allowed $(m,T)$ in the proxy $\eta=3$ shell, obtained using Eqs. (\ref{eq.prox2})-(\ref{eq.prox6}) are listed in Table 1.

\begin{table}
\begin{center}
\caption{Leading $SU(3)$ irrep $(\lambda_H , \mu_H)$ for a given number $m$ of 
nucleons and isospin $T=|T_Z|$. Results are  shown for $\eta=3$ shell with $m \geq 3$. The irreps are shown in the table as $(\lambda_H ,\mu_H)^{m,T}$. For $m$ odd, $2T$ instead of $T$ is shown.}
{\tiny{
\begin{tabular}{l}
\hline
$\eta=3:\;\;$m$\;\;$even \\
$(12, 0)^{ 4, 0}$,$(10, 1)^{ 4, 1}$,$( 8, 2)^{ 4, 2}$,$(14, 2)^{ 6, 0}$,
$(14,2)^{ 6, 1}$,$(13, 1)^{ 6, 2}$,
$(12, 0)^{ 6, 3}$,$(16, 4)^{ 8, 0}$,$(17, 2)^{ 8, 1}$,\\
$(18, 0)^{ 8, 2}$,$(14, 2)^{ 8, 3}$,$(10, 4)^{ 8, 4}$,
$(20, 2)^{10, 0}$,$(20, 2)^{10, 1}$,$(18, 3)^{10, 2}$,$(16, 4)^{10, 3}$,
$(13, 4)^{10, 4}$,$(10, 4)^{10, 5}$,\\
$(24, 0)^{12, 0}$,$(21, 3)^{12, 1}$,$(18, 6)^{12, 2}$,$(17, 5)^{12, 3}$,
$(16, 4)^{12, 4}$,$(14, 2)^{12, 5}$,
$(12, 0)^{12, 6}$,$(22, 4)^{14, 0}$,$(22, 4)^{14, 1}$,\\
$(20, 5)^{14, 2}$,$(18, 6)^{14, 3}$,$(18, 3)^{14, 4}$,
$(18, 0)^{14, 5}$,$(12, 3)^{14, 6}$,$( 6, 6)^{14, 7}$,$(20, 8)^{16, 0}$,
$(21, 6)^{16, 1}$,$(22, 4)^{16, 2}$,\\
$(21, 3)^{16, 3}$,$(20, 2)^{16, 4}$,$(16, 4)^{16, 5}$,$(12, 6)^{16, 6}$,
$( 7, 7)^{16, 7}$,$( 2, 8)^{16, 8}$,
$(20, 8)^{18, 0}$,$(20, 8)^{18, 1}$,$(22, 4)^{18, 2}$,\\
$(24, 0)^{18, 3}$, $(19, 4)^{18, 4}$,$(14, 8)^{18, 5}$,$(11, 8)^{18, 6}$,
$( 8, 8)^{18, 7}$,$( 4, 7)^{18, 8}$,$( 0, 6)^{18, 9}$,
$(20, 8)^{20, 0}$,$(21, 6)^{20, 1}$,\\
$(22, 4)^{20, 2}$,$(20, 5)^{20, 3}$,$(18, 6)^{20, 4}$,$(14, 8)^{20, 5}$,
$(10,10)^{20, 6}$,$( 8, 8)^{20, 7}$,
$( 6, 6)^{20, 8}$,$( 3, 3)^{20, 9}$,$( 0, 0)^{20,10}$,\\
$(22, 4)^{22, 0}$,$(22, 4)^{22, 1}$,$(19, 7)^{22, 2}$,
$(16,10)^{22, 3}$,$(15, 9)^{22, 4}$,$(14, 8)^{22, 5}$,$(11, 8)^{22, 6}$,
$( 8, 8)^{22, 7}$,$( 7, 4)^{22, 8}$,\\
$( 6, 0)^{22, 9}$,$(24, 0)^{24, 0}$,$(20, 5)^{24, 1}$,$(16,10)^{24, 2}$,
$(14,11)^{24, 3}$,$(12,12)^{24, 4}$,
$(12, 9)^{24, 5}$,$(12, 6)^{24, 6}$,$(10, 4)^{24, 7}$,\\
$( 8, 2)^{24, 8}$,$(18, 6)^{26, 0}$,$(18, 6)^{26, 1}$,
$(15, 9)^{26, 2}$,$(12,12)^{26, 3}$,$(11,11)^{26, 4}$,$(10,10)^{26, 5}$,
$(11, 5)^{26, 6}$,$(12, 0)^{26, 7}$,\\
$(12,12)^{28, 0}$,$(13,10)^{28, 1}$,$(14, 8)^{28, 2}$,$(12, 9)^{28, 3}$,
$(10,10)^{28, 4}$,$(10, 7)^{28, 5}$,
$(10, 4)^{28, 6}$,$( 8,14)^{30, 0}$,$( 8,14)^{30, 1}$,\\
$(10,10)^{30, 2}$,$(12, 6)^{30, 3}$,$(11, 5)^{30, 4}$,
$(10, 4)^{30, 5}$,$( 4,16)^{32, 0}$,$( 5,14)^{32, 1}$,$( 6,12)^{32, 2}$,
$( 9, 6)^{32, 3}$,$(12, 0)^{32, 4}$,\\
$( 2,14)^{34, 0}$,$( 2,14)^{34, 1}$,$( 4,10)^{34, 2}$,$( 6, 6)^{34, 3}$,
$( 0,12)^{36, 0}$,$( 1,10)^{36, 1}$,
$( 2, 8)^{36, 2}$,$( 0, 6)^{38, 0}$,$( 0, 6)^{38, 1}$,$( 0, 0)^{40, 0}$\\
\hline
\hline
$\eta=3:\;\;$ m$\;\;$odd \\
$( 9, 0)^{ 3, 1}$, $( 7, 1)^{ 3, 3}$, $(13, 1)^{ 5, 1}$, $(11, 2)^{ 5, 3}$, $(10, 1)^{ 5, 5}$, $(15, 3)^{ 7, 1}$, $(16, 1)^{ 7, 3}$, $(15, 0)^{ 7, 5}$, \\
$(11, 2)^{ 7, 7}$, $(18, 3)^{ 9, 1}$, $(19, 1)^{ 9, 3}$, $(17, 2)^{ 9, 5}$, $(13, 4)^{ 9, 7}$, $(10, 4)^{ 9, 9}$, $(22, 1)^{11, 1}$, $(19, 4)^{11, 3}$, \\
$(17, 5)^{11, 5}$, $(16, 4)^{11, 7}$, $(13, 4)^{11, 9}$, $(11, 2)^{11,11}$, $(23, 2)^{13, 1}$, $(20, 5)^{13, 3}$, $(18, 6)^{13, 5}$, $(17, 5)^{13, 7}$, \\
$(17, 2)^{13, 9}$, $(15, 0)^{13,11}$, $( 9, 3)^{13,13}$, $(21, 6)^{15, 1}$, $(22, 4)^{15, 3}$, $(20, 5)^{15, 5}$, $(19, 4)^{15, 7}$, $(19, 1)^{15, 9}$, \\
$(15, 3)^{15,11}$, $( 9, 6)^{15,13}$, $( 4, 7)^{15,15}$, $(20, 8)^{17, 1}$, $(21, 6)^{17, 3}$, $(23, 2)^{17, 5}$, $(22, 1)^{17, 7}$, $(17, 5)^{17, 9}$, \\
$(13, 7)^{17,11}$, $(10, 7)^{17,13}$, $( 5, 8)^{17,15}$, $( 1, 7)^{17,17}$, $(20, 8)^{19, 1}$, $(21, 6)^{19, 3}$, $(23, 2)^{19, 5}$, $(21, 3)^{19, 7}$, \\
$(16, 7)^{19, 9}$, $(12, 9)^{19,11}$, $( 9, 9)^{19,13}$, $( 7, 7)^{19,15}$, $( 3, 6)^{19,17}$, $( 0, 3)^{19,19}$, $(21, 6)^{21, 1}$, $(22, 4)^{21, 3}$, \\
$(19, 7)^{21, 5}$, $(17, 8)^{21, 7}$, $(16, 7)^{21, 9}$, $(12, 9)^{21,11}$, $( 9, 9)^{21,13}$, $( 7, 7)^{21,15}$, $( 6, 3)^{21,17}$, $( 3, 0)^{21,19}$, \\
$(23, 2)^{23, 1}$, $(19, 7)^{23, 3}$, $(16,10)^{23, 5}$, $(14,11)^{23, 7}$, $(13,10)^{23, 9}$, $(13, 7)^{23,11}$, $(10, 7)^{23,13}$, $( 8, 5)^{23,15}$, \\ 
$( 7, 1)^{23,17}$, $(21, 3)^{25, 1}$, $(17, 8)^{25, 3}$, $(14,11)^{25, 5}$, $(12,12)^{25, 7}$, $(11,11)^{25, 9}$, $(11, 8)^{25,11}$, $(12, 3)^{25,13}$, \\
$(10, 1)^{25,15}$, $(15, 9)^{27, 1}$, $(16, 7)^{27, 3}$, $(13,10)^{27, 5}$, $(11,11)^{27, 7}$, $(10,10)^{27, 9}$, $(10, 7)^{27,11}$, $(11, 2)^{27,13}$, \\
$(10,13)^{29, 1}$, $(11,11)^{29, 3}$, $(13, 7)^{29, 5}$, $(11, 8)^{29, 7}$, $(10, 7)^{29, 9}$, $(10, 4)^{29,11}$, $( 6,15)^{31, 1}$, $( 7,13)^{31, 3}$, \\
$( 9, 9)^{31, 5}$, $(12, 3)^{31, 7}$, $(11, 2)^{31, 9}$, $( 3,15)^{33, 1}$, $( 4,13)^{33, 3}$, $( 6, 9)^{33, 5}$, $( 9, 3)^{33, 7}$, $( 1,13)^{35, 1}$, \\
$( 2,11)^{35, 3}$, $( 4, 7)^{35, 5}$, $( 0, 9)^{37, 1}$, $( 1, 7)^{37, 3}$, $( 0, 3)^{39, 1}$ \\
\hline
\end{tabular}
}}
\end{center}
\end{table}
Given a $(\lambda_H,\mu_H)$ irrep, the nucleus will be prolate if
$\lambda_H > \mu_H$ and oblate for $\lambda_H < \mu_H$ \cite{beta-gam}. Significantly, it is easy to see from the irreps in Table 1, that the onset of oblate shapes will be much later than the mid-shell, i.e. we need $m >> 20$ and the oblate shape starts only from $m=29$. An interesting example is $^{72}$Kr. For this
nucleus, ground state is oblate if we consider the valence nucleons to be only in $^2p_{3/2}$, $^1f_{5/2}$ and $^2p_{1/2}$ orbits with $^{56}$Ni core (then we have $\eta=2$ shell with pseudo $SU(3)$). However, if the proxy-$SU(3)$ model in conjunction with the proxy-$SU_{ST}(4)$ is used, then $\eta=3$ shell with results in Table 1 will apply for identifying the lowest $SU(3)$ irrep and there by the shape. This gives prolate shape for $^{72}$Kr. Experimental data for this nucleus indicates prolate with some oblate mixing; see for example \cite{KS,Kr72}. Thus, proxy-$SU(3)-SU_{ST}(4)$ classification is useful for this nucleus and in general for A=60-90 nuclei.  

Going further, it is indeed possible to identify all the allowed $\{f\}$ of $U(10)$ with $m$ changing from 0 to 40 and for each of this $\{f\}$ obtain the full $\{f\} \rightarrow (\lambda,\mu)$ reductions. Methods for these are described in \cite{su3-book} and the tabulations are available in \cite{Kota-tab}. As an example, for $m=12$, $14$ and $16$ some of the low-lying $SU(3)$ irreps are listed for some lowest $\{f\}$'s in Table 2. These will be useful in detailed calculation that include mixing of $SU(3)$ irreps (also those of $SU_{ST}(4)$).

\begin{table}
\begin{center}
\caption{Lowest few $SU(3)$ irreps $(\lambda , \mu)$ for some irreps $\{f\}$ of $U(10)$ for number of nucleons $m=12$,14 and 16. In the table, The $r$ in $(\lambda , \mu)^r$  irreps gives the multiplicity of the $(\lambda, \mu)$ irrep and when $r$ is not shown implies $r=1$.}
\begin{tabular}{cl}
\hline
$\{f\}\;\;\;\;$ & $(\lambda,\mu)^r$ \\
$\{4^3\}$ & (24,0), (21,3), (18,6), (19,4), (20,2)$^3$, $\ldots$ \\
$\{4^231\}$ & (21,3), (22,1), (18,6), (19,4)$^3$, (20,2)$^4$, (21,0), $\ldots$ \\
$\{4^32\}$ & (22,4), (24,0), (19,7), (20,5)$^2$, (21,3)$^3$, (22,1)$^2$, $\ldots$ \\
$\{4^31^2\}$ & (23,2), (20,5)$^2$, (21,3)$^2$, (22,1)$^2$, $\ldots$ \\
$\{4^4\}$ & (20,8), (22,4), (24,0), (18,9), (19,7)$^2$, (20,5)$^3$, (21,3)$^3$, (22,1), $\ldots$ \\
$\{4^331\}$ & (21,6), (22,4), (23,2), (18,9)$^2$, (19,7)$^4$, (20,5)$^7$, (21,3)$^7$, (22,1)$^5$, $\ldots$ \\
\hline
\end{tabular}
\end{center}
\end{table}
 
\section{Evidence for proxy-$SU(4)$ from spectroscopic results for some select nuclei in A=60-90 region}

In this Section we will present some results that give indirect evidence for the application of proxy-$SU_{ST}(4)$ symmetry.
  
\subsection{Triple forking in $^{68}$Ge}
 
Germanium isotopes are in transitional region and for example in $^{68}$Ge, the ground state rotational band
trifurcates as energy increases exhibiting three close lying $8^+$ states (’triple forking’), around 5 MeV excitation, with roughly the same $E2$ strength to the lone $6^+$ state of the ground band. In a pseudo-$SU(3)$ description used is $^{56}$Ni core and distributing the 12 valence nucleons in $r3$ orbits where ($r3$ = $^1p_{3/2}$, $^1p_{1/2}$, $^0f_{5/2}$) and $^0g_{9/2}$ orbit giving configurations
$(r3)^{m_{na}} (^0g_{9/2})^{m_{in}}$ with $m=12=m_{na} +m_{in}$ \cite{psu3}. In the $r3$ space we have pseudo-$SU(3)\otimes SU_{ST}(4)$ symmetry and in the $^0g_{9/2}$ space the proton-neutron $SO(5)$ pairing algebra. In the calculations in \cite{psu3},chosen are $m_{na}=12$, $10$ and $8$ (then $m_{in}=0$, $2$ and $4$ respectively) along with a few $SU_{ST}(4)$ and
$SU(3)$ irreps plus a few seniority states in $^0g_{9/2}$ space.
The pseudo-$SU(3)$ calculations describe quite well the observed triple forking with three close lying $8^+$ levels and their $E2$ decay strengths. In another set of calculations,
deformed shell model (DSM) based on Hartree-Fock single particle states (see \cite{KS} for details regarding DSM) was employed \cite{KS-8} and they also brought out the triple-forking feature in $^{68}$Ge. In DSM calculations, sp orbits $^1p_{3/2}$, $^1p_{1/2}$, $^0f_{5/2}$ and $^0g_{9/2}$ with sp energies 0.0, 0.78, 1.08 and 5.0 MeV respectively are used and the realistic Kuo interaction is employed. It is seen that the first $8^+_1$ state around 5MeV excitation arises from prolate two neutron aligned deformed intrinsic state with structure $(r3)^{10}(g_{9/2})^2$ and $8_2$ that is close to the $8^+_1$ is a mixed state dominated by a oblate neutron aligned intrinsic state. Finally, the other close lying $8^+_3$ state is dominated by $(r3)^{12}$ structure \cite{KS-8}. 
Both the pseudo-$SU(3)$ and DSM descriptions show that for $^{68}$Ge it is essential to consider all the four $^1p_{3/2}$, $^1p_{1/2}$, $^0f_{5/2}$ and $^0g_{9/2}$ orbits. Let us add that DSM studies also showed that these orbits are needed to describe
shape coexistence and shape changes seen in Se isotopes \cite{KS}.
Therefore, proxy-$SU(3)-SU_{ST}(4)$ scheme given by Eq. (\ref{eq.prox1}) is expected to be useful for Ge and Se isotopes.
 
\subsection{Spectroscopy of $^{66}$As}

Experimental data for $^{66}$As consists of the ground $T=1$ band
and four excited $T=0$ bands. Shell model (SM) describes these bands \cite{PSK-as66} and the structure of the lowest three more clearly brought out by DSM while that of the fourth band by cranked Nilsson–Strutinsky calculations \cite{Aberg}. In these
studies again the sp orbits $^1p_{3/2}$, $^1p_{1/2}$, $^0f_{5/2}$ and $^0g_{9/2}$ are used. In SM and DSM studies the jj44b interaction (with sp energies) developed recently by Brown and Lisetskiy is employed. The SM calculations with jj44b interaction describe the observed $T=1$ band (band 1), isobaric analogue of $^{66}$Ge ground band, reasonably well and predicted a structural change at $8^+$. Using DSM it is seen that at $J^\pi = 8^+$ there is a band crossing due to the occupancy of a proton and a neutron in $^0g_{9/2}$ orbit. Also, shell model gives the first $T=0$ band (band 2) above the $T=1$ band. For the third $9^+$ band with $T=0$, the shell model energies, B(E2)'s and quadrupole moments are
consistent with the interpretation in terms of an aligned isoscalar $np$ pair in $^0g_{9/2}$ orbit coupled to the $^{64}$Ge ground band. Similarly in band 4, the $9^+$ level and a
close-lying $5^+$ level are found to be isomeric states. Finally, the shell model energies for band 5 members, show that the observed levels are most likely negative parity levels. All this description of the 5 bands in $^{66}$As again need the four $^1p_{3/2}$, $^1p_{1/2}$, $^0f_{5/2}$ and $^0g_{9/2}$ orbits and hence the proxy-$SU(3)-SU_{ST}(4)$ scheme is expected to be useful for this nucleus. 

\subsection{Ground state masses}

Janecke et.al. \cite{Ja-PRC,Ja-PLB} investigated using experimentally determined masses of nuclei, the differences in the excitation energies $\Delta_{T^\prime T}(A)$ between isobaric analog states with isospins $T^\prime$ and $T$ in the same nucleus. These states are ground states or states analog to ground states in neighboring isobars. Focusing on N=Z nuclei, it is seen that $\Delta_{T^\prime T}(A)$ is sensitive to the symmetry energy ($sym$). More importantly, it will allow us to deduce the value of $x$ in $E_{sym} = \{a(A,T\}/A\}\,T(T+x)$. Note that with only isospin symmetry we have $x=1$ and with quarteting or $SU_{ST}(4)$ symmetry, we have $x=4$. In particular, the ratio of $\Delta_{2,0}(A)$ and average of $\Delta_{1,0}(A)$ depends only on $x$ and it will be independent
of the pairing energy; note that $\lan \Delta_{1,0}(A)\ran = [\Delta_{1,0}(A-2) + 2\Delta_{1,0}(A) +\Delta_{1,0}(A+2)]/4$.
In fact, it is seen that,
\be
R(A) = \dis\f{\Delta_{2,0}(A)}{\lan \Delta_{1,0}(A) \ran} = \dis\f{4+2x}{1+x}\;.
\label{eq.prox7}
\ee
Then, $R(A)=3$ for isospin symmetry ($x=1$) and $2.4$ for $SU(4)$
symmetry ($x=4$). Values of $R(A)$ calculated using experimental masses clearly showed that $R(A) \sim 2.4$ for N=Z nuclei around $A=80$. For example for $A$ in the range 80-88, it is found that
$R = 2.45 \pm 0.08$ \cite{Ja-PRC}. Thus $SU(4)$ symmetry is expected to be important at least for nuclei around $A=80$. The analysis of $R$ values can be improved using the latest mass compilations given in \cite{Mass-1,Mass-2} and they are unlikely to change the conclusion that $SU(4)$ symmetry is important in A=60-90 region. 

\subsection{GT distributions}

Gamow-Teller (GT) distributions provide a test of $SU_{ST}(4)$ symmetry as the GT operator is a generator of $SU_{ST}(4)$. However, in testing proxy-$SU_{ST}(4)$ using GT distributions, it
is necessary to note that the $X^{11}_{\mu,\nu}$ generator given by Eq. (3) is only a proxy-GT operator. Therefore, a renormalization of the $X$-operator is needed. Thus, an effective GT operator has to be employed in the calculations that use proxy $\eta=3$ shell. There is recent data on GT distributions in $^{70}$Kr $\rightarrow$ $^{70}$Br \cite{KrBr}. There are results obtained using proton-neutron quasi-particle random-phase approximation (pnQRPA) with prolate and oblate shapes \cite{Surg}. The pnQRPA results show that in order to describe the data, one needs the full $(r3)-( ^0g_{9/2})$ space. Clearly, GT distribution data give new insights into shape changes and shape coexistence \cite{Lenzi}. As mentioned in the introduction,
pseudo-$SU(3)\otimes SU(4)$ in $(r3)$ space with pn-pairing $SO(5)$ in $( ^0g_{9/2})$ may also describe data \cite{psu4}.

\subsection{Summary}

Results discussed in Sections 3.1-3.4 show that
the sp orbits $(r3)-(^0g_{9/2})$ are essential for $A=60-90$ nuclei and hence the need for proxy-$SU_{ST}(4)$ schemes for 
$A=60-90$ nuclei [with $(r3)$ orbits and $^0g_{9/2}$ orbit all active, the $SU(4)$ algebra one can construct is only the proxy-$SU(4)$ algebra]. With this evidence, it is important to explore further the symmetry limits of $U(40) \supset U(10) \otimes SU_{ST}(4)$ and this is the topic of the next two Sections. 

\section{New proxy-$SO(6)$, proxy-$SU(5)$ and proxy-$SO(10)$ schemes}

With proxy-$SU(4)$ symmetry, the SGA in $(r3)-( ^0g_{9/2})$ space is $U(40) \supset U(10) \otimes SU_{ST}(4)$ as already mentioned in
Section 2. Then, we have proxy-$pf$ ($\ell=1$ and $3$) orbitals giving the spatial $U(10)$ algebra. The $U(10)$ contains not only the $SU(3)$ algebra described
in Section 2 but also $SO(6)$, $SU(5)$ ans $SO(10)$ algebras (these $U(10)$ subalgebras are first mentioned in \cite{Hecht-u10} and used in IBFM studies in \cite{Kota-ibfm}). {\it A new element that was not recognized earlier is that, just as the case with $SU(3)$, the $SO(6)$, $SU(5)$ and $SO(10)$ algebras appear twice
as shown below}. 

Firstly,
the $SU(4) \sim SO(6)$ subalgebra arises from the recognition that two particle symmetric states with each particle carrying angular momentum $j=3/2$ will generate $\ell=1,3$. Also, $SO(6)$ contains $SO(5)$ subalgebra. Then,
\be
SO(6)\; limit\;:\;\;\;
\{1\}_{U(10)} \rightarrow \l[\{2\}_{SU(4)} \sim \l[1,1,1\r]_{SO(6)}\r] \rightarrow \l[1,1\r]_{SO(5)} \rightarrow \{\ell=1,3\}_{SO(3)}\;.
\label{eq.prox8}
\ee
Note that $SO(6) \supset SO(5)$ is same as $SU(4) \supset Sp(4)$. 
Similarly, a $SU(5)$ subalgebra arises from the recognition that two particle antisymmetric states with each particle carrying angular momentum $j=2$ will generate $\ell=1,3$. Also, $SU(5)$ contains $SO(5)$ subalgebra. Then,
\be
SU(5)\; limit\;:\;\;\;
\{1\}_{U(10)} \rightarrow \{1^2\}_{SU(5)} \rightarrow \l[1,1\r]_{SO(5)} \rightarrow \{\ell=1,3\}_{SO(3)}\;.
\label{eq.prox9}
\ee
Whenever needed, in the following we use $U(4)$ and $U(5)$ instead of $SU(4)$ and $SU(5)$. Besides these limits, $U(10)$ also admits isoscalar plus isovector pairing algebra $SO(10)$ \cite{Flow,Pang,KoCas}. Then we have,
\be
SU(10)\; limit\;:\;\;\;
\{1\}_{U(10)} \rightarrow \{1\}_{SO(10)} \rightarrow \l[1,1\r]_{SO(5)} \rightarrow \{\ell=1,3\}_{SO(3)}\;.
\label{eq.prox10}
\ee
In the following subsections we will give the generators and quadratic Casimir operators for the various algebras in Eqs. (\ref{eq.prox8})-(\ref{eq.prox10}). Later, in Section 5 we will consider the reductions of the various irreps that will allow
one to construct typical spectra in the three limits.

\subsection{Proxy-$SO(6)$ limit: generators and quadratic Casimir operators}

In the $SO(6)$ limit the group chain and the irrep labels are
\be
SO(6)\;:\;\;\;\l.\l|\barr{ccccccc} U(10) & \supset & SO(6) & \supset & SO(5) & \supset & SO(3) \\ \l\{f\r\} & & \l[\sigma_1,\sigma_2,\sigma_3\r] & & \l[v_1,v_2\r] && L \earr \r.\ran\;.
\label{eq.prox11}
\ee
Note that as an intermediate we often use $U(4)$ with its irreps $\{\cf\} = \{\cf_1,\cf_2,\cf_3,\cf_4\}$ and $U(4) \supset SU(4)$ with $SU(4)$ isomorphic to $SO(6)$ in Eq. (\ref{eq.prox11}); $SU(4)$ here should not be confused with $SU_{ST}(4)$ as one is in orbital space and other in spin-isospin space. Importantly, $\{\cf\}$ is related in a simple manner to $\l[\sigma_1,\sigma_2,\sigma_3\r]$,
\be
\l[\sigma_1,\sigma_2,\sigma_3\r] = \l[ \dis\f{\cf_1+\cf_2-\cf_3-\cf_4}{2}, \dis\f{\cf_1 -\cf_2+\cf_3-\cf_4}{2}, \dis\f{\cf_1-\cf_2-\cf_3+\cf_4}{2}\r]\;.
\label{eq.prox11a}
\ee
Following the results derived for sdgIBM in the past \cite{sdg-1,sdg-2} and using the fact that $U(4)$ (equivalently $SO(6)$) arises from symmetric coupling of two $j=3/2$ particles give immediately the generators $g^{L_0=0,1,2,3}_q$ of the subalgebras in Eq. (\ref{eq.prox11}) [generators of $U(10)$ are given by Eq. (\ref{eq.proxb})]. These are
$g^{L_{12}}_q$ where
\be
\barr{l}
g^{L_{12}}_q(\alpha) = 2\,\dis\sum_{\ell_1,\ell_2=1,3} A_{\ell_1,\ell_2}(\alpha) \,\dis\sqrt{(2\ell_1+1)(2\ell_2+1)}\l\{\barr{ccc} \ell_1 & \ell_2 & L_{12} \\ \f{3}{2} & \f{3}{2} & \f{3}{2} \earr \r\}\; \l( a^\dagger_{\ell_1} \tilde{a}_{\ell_2}\r)^{L_{12},0,0}_q\;;\\
A_{\ell_1,\ell_2}(\alpha) = \l\{\delta_{\ell_1 \ell_2} + \alpha\,(1-\delta_{\ell_1 \ell_2})\r\}\;\;\mbox{with}\;\; \alpha = +1\;\;\;\mbox{or}\;\;\;-1\;.
\earr \label{eq.prox12}
\ee
With this, the generators of $U(4)$ are $\{g^0_q(\alpha), g^1_q(\alpha), g^2_q(\alpha), g^3_q(\alpha)\}$ and those of $SU(4)$ or $SO(6)$ are $\{g^1_q(\alpha), g^2_q(\alpha), g^3_q(\alpha)\}$. Similarly, the generators of $SO(5)$ and $SO(3)$ are $\{g^1_q(\alpha), g^3_q(\alpha)\}$ and $\{g^1_q(\alpha)\}$ respectively; note that $g^0$ and $g^1$ do not depend on $\alpha$ and $L^1_q=\sqrt{20}\,g^1_q$. It is important to emphasize that there will be two $SO^{\alpha}(6)$ algebraic limits and they correspond to $\alpha=1$ or $-1$ in Eq. (\ref{eq.prox12}). This is similar to the two $SU(3)$ algebras mentioned in Section 2. Given the generators, the quadratic Casimir invariants $\cc_2$ and their eigenvalues over the respective irreps are \cite{Kota-ibfm,Wy-70,Wy-74},
\be
\barr{l}
\cc_2(U(10)) = 4 \dis\sum_{\ell_1,\ell_2=1,3; L} u^{L,0,0}(\ell_1,\ell_2) \cdot u^{L.0.0}(\ell_2,\ell_1)\;,\\
\cc_2(SO^{\alpha}(6)) = 4 \dis\sum_{k=1}^3 g^k(\alpha) \cdot g^k(\alpha)\;,\\
\cc_2(SO^{\alpha}(5)) = 4 \dis\sum_{k=1,3}  g^k(\alpha) \cdot g^k(\alpha)\;,\\
\cc_2(SO(3)) = 20 g^1 \cdot g^1\;,\\
\lan \cc_2(U(10)) \ran^{\{f\}} = \dis\sum_{i=1}^{10} f_i(f_i+11-2i)\;,\\
\lan \cc_2(SO^{\alpha}(6))\ran^{\l[\sigma_1,\sigma_2,\sigma_3\r]}
= \l[\sigma_1(\sigma_1+4) + \sigma_2(\sigma_2+2) + \sigma_3^3\r]\;,\\
\lan \cc_2(SO^{\alpha}(5))\ran^{\l[v_1,v_2\r]}
= \l[v_1(v_1+3) + v_2(v_2+1) \r]\;,\\
\lan \cc_2(SO(3))\ran^L = L(L+1)\;.
\earr \label{eq.prox13}
\ee
It is important to notice from Eq. (\ref{eq.prox13}) that the eigenvalues of the Casimir invariants do not depend on $\alpha$
and therefore, the $SO^{\alpha}(6)$ symmetry limit generates same spectra for both $\alpha=+1$ and $-1$. However, the electromagnetic decay properties and particle transfer strength will be different as discussed in \cite{Kota-1,Kota-2}. 
 
\subsection{Proxy-$SU(5)$ limit: generators and quadratic Casimir operators}

In the $SU(5)$ limit the group chain and the irrep labels are
\be
SU(5)\;:\;\;\;\l.\l|\barr{ccccccc} U(10) & \supset & SU(5) & \supset & SO(5) & \supset & SO(3) \\ \l\{f\r\} & & \l\{a\r\} & & \l[v_1,v_2\r] && L \earr \r.\ran\;.
\label{eq.prox14}
\ee
Note that as an intermediate we often use $U(5)$ and its irreps $\{\ca\} = \{\ca_1,\ca_2,\ca_3,\ca_4,\ca_5\}$ and $U(5) \supset SU(5)$ with 
\be
\{\ca\}_{U(5)} \rightarrow \{a\}_{SU(5)} = \l\{a_1,a_2,a_3,a_4\r\} = \l\{\ca_1-\ca_5,\ca_2-\ca_5, \ca_3-\ca_5, \ca_4-\ca_5\r\}\;.
\label{eq.prox15}
\ee
As before, following the results derived for sdgIBM in the past \cite{sdg-1,sdg-2} and using the fact that $U(5)$ arises from antisymmetric coupling of two $j=2$ particles give immediately the generators $h^{L_0=0,1,2,3,4}_q$ of the subalgebras in Eq. (\ref{eq.prox14}). These are,
\be
\barr{l}
h^{L_{12}}_q(\alpha) = 2\,\dis\sum_{\ell_1,\ell_2=1,3} A_{\ell_1,\ell_2}(\alpha) \,\dis\sqrt{(2\ell_1+1)(2\ell_2+1)}\l\{\barr{ccc} \ell_1 & \ell_2 & L_{12} \\ 2 & 2 & 2 \earr \r\}\; \l( a^\dagger_{\ell_1} \tilde{a}_{\ell_2}\r)^{L_{12},0,0}_q\;;\\
A_{\ell_1,\ell_2}(\alpha) = \l\{\delta_{\ell_1 \ell_2} + \alpha\,(1-\delta_{\ell_1 \ell_2})\r\}\;\;\mbox{with}\;\; \alpha = +1\;\;\;\mbox{or}\;\;\;-1\;.
\earr \label{eq.prox16}
\ee
With this, the generators of $U(5)$ are $\{h^0_q(\alpha), h^1_q(\alpha), h^2_q(\alpha), h^3_q(\alpha), h^4_q(\alpha) \}$ and those of $SU(5)$ are $\{h^1_q(\alpha), h^2_q(\alpha), h^3_q(\alpha), h^4_q(\alpha)\}$. Similarly, the generators of $SO(5)$ and $SO(3)$ are $\{h^1_q(\alpha), h^3_q(\alpha)\}$ and $\{h^1_q(\alpha)\}$ respectively; note that $h^0$ and $h^1$ do not depend on $\alpha$ and, for $L_{12}=1,3$ 
$$
g^{L_{12}}_q(\alpha)= \sqrt{2}\;h^{L_{12}}_q(\alpha)
$$
within a overall sign factor. It is important to emphasize that there will be two $SU^{\alpha}(5)$ algebraic limits and they correspond to $\alpha=1$ or $-1$ in Eq. (\ref{eq.prox16}). This is similar to the two $SU(4)$ algebras mentioned in Section 4.1. Given the generators, the quadratic Casimir invariants $\cc_2$ and their eigenvalues over the respective irreps are \cite{Kota-ibfm,Wy-70,Wy-74},
\be
\barr{l}
\cc_2(SU^{\alpha}(5)) = 4 \dis\sum_{k=1}^4 h^k(\alpha) \cdot h^k(\alpha)\;,\\
\cc_2(SO^{\alpha}(5)) = 8 \dis\sum_{k=1,3}  h^k(\alpha) \cdot h^k(\alpha)\;,\\
\lan \cc_2(SU^{\alpha}(5))\ran^{\{a\}}
= \dis\sum_{i=1}^4 a_i(a_i+6-2i) -\f{1}{5} \l\{\sum_{i=1}^4 a_i\r\}^2\\
\lan \cc_2(SO^{\alpha}(5))\ran^{\l[v_1,v_2\r]}
= \l[v_1(v_1+3) + v_2(v_2+1) \r]\;.
\earr \label{eq.prox17}
\ee
The $\cc_2$ of $U(10)$ and $SO(3)$ are already given in Eq. (\ref{eq.prox13}). It is important to notice from Eq. (\ref{eq.prox13}) that the eigenvalues of the Casimir invariants do not depend on $\alpha$
and therefore, the $SU^{\alpha}(5)$ symmetry limit generates same spectra for both $\alpha=+1$ and $-1$. However, the electromagnetic decay properties and particle transfer strength will be different. 

\subsection{Proxy-$SO(10)$ limit: generators and quadratic Casimir operators}

In the $SO(10)$ limit the group chain and the irrep labels are
\be
SO(10)\;:\;\;\;\l.\l|\barr{ccccccc} U(10) & \supset & SO(10) & \supset & SO(5) & \supset & SO(3) \\ \l\{f\r\} & & \l[\omega\r] & & \l[v_1,v_2\r] && L \earr \r.\ran\;.
\label{eq.prox18}
\ee
Note that $\l[\omega\r]$ will be a five row irrep,
\be
\l[\omega\r] = \l[\omega_1,\omega_2,\omega_3,\omega_4,\omega_5\r]\;;\;\;\;\;
\omega_1 \ge \omega_2 \ge \omega_3 \ge \omega_4 \ge \l|\omega_5\r| \ge 0\;. 
\label{eq.prox19}
\ee
Thus, $\omega_5$ above can be positive or negative. 
It is well known that the $SO(10)$ limit corresponds to equal isoscalar plus isovector pairing and it is possible to define seniority quantum number $v$ given a $\l[\omega\r]$ irrep \cite{Flow,Pang,KoCas}. The generators of $SO(10)$ are given for example in \cite{KoCas} and they are
\be
\barr{l}
V^L_q(\ell,\ell) = u^{L,0,0}_q(\ell,\ell)\;;\;\;\;\ell=1,3\;\;\;\mbox{and}\;\;\;\,L\;\mbox{odd}\;,\\
V^L_q(3,1;\beta) = \l[\beta\,(-1)^L\r]^{1/2}\,\l\{u^{L,0,0}_q(3,1) + \beta\, (-1)^L u^{L,0,0}_q(1,3)\r\}\;;\;\;\;
\beta=+1\;\mbox{or} -1\;.
\earr \label{eq.prox20}
\ee
Thus, there will be two $SO^{\beta}(10)$ algebras that correspond to $\beta=+1$ or $-1$. Comparing the generators in Eq. (\ref{eq.prox20}) with the $SO(5)$ generators in Sections 4.1 and 4.2, it is seen that the $SO(5)$ sublgebra of $SO(10)$ in Eq. (\ref{eq.prox18}) will be same as the $SO(5)$ algebras in $SO(6)$ and $SU(5)$ limits only when $\beta=-1$ in Eq. (\ref{eq.prox20}), i.e. for $SO^{(-)}(10)$. The quadratic Casimir operator of $SO(10)$ and its eigenvalues are,
\be
\barr{l}
\cc_2(SO^{\beta}(10)) = 8 \dis\sum_{\ell=1,3;\;L\;odd} V^L(\ell,\ell)
\cdot V^L(\ell,\ell) + 4 \dis\sum_{L} V^L(3,1;\beta) \cdot
V^L(3,1;\beta)\;,\\
\lan \cc_2(SO^{\beta}(10)) \ran^{\l[\omega\r]} = \dis\sum_{i=1}^5 
\omega_i (\omega_i+10-2i)\;.
\earr  \label{eq.prox21}
\ee
Note that the eigenvalues do not depend on $\beta$.

\section{Irrep reductions in $SO(6)$, $SU(5)$ and $SO(10)$ limits and shapes with the symmetry limits}

In order to construct the typical spectra generated by the symmetry limits and to derive various other properties generated by them, including determination of the ground state structure for example, we need to first determine the labels in Eqs. (\ref{eq.prox11}), (\ref{eq.prox14}) and (\ref{eq.prox18}), i.e. the irrep reductions for a given $U(10)$ irrep. The rules for these are in general known \cite{Wy-70,Little} but in practice they are quite complex. We will describe the results for the $SO(6)$ limit in some detail and other two briefly. 
Before proceeding further, it is important to add that the reductions in the symmetry limits are also considered in
the context of $2p-1f$ shell nuclei in \cite{Hecht-u10}. However,
the results in \cite{Hecht-u10} are restricted to two column irreps of $U(10)$ as the focus there is on identical nucleon systems. In the present work, as we consider proxy-$SU(4)$, the
interest is in four column irreps of $U(10)$ and the results for these are given in Sections 5.1-5.3. In Section 5.4 we will discuss briefly the ground state
structure generated by the symmetry limits. 

\subsection{Proxy-$SO(6)$ limit: irrep reductions}

Given a symmetric irrep $\{n\}$ of $U(10)$, the irreps $\{\cf\}$ of $U(4)$ follow easily from \cite{Little} as $\{1\}_{U(10)} \rightarrow \{2\}_{U(4)}$ giving,
\be
\barr{l}
\{n\}_{U(10)} = \dis\sum \{2n_1, 2n_2, 2n_3, 2n_4\}_{U(4)} \oplus\;;\\
n_1 \ge n_2 \ge n_3 \ge n_4 \ge 0\;,\;\;\;n=n_1+n_2+n_3+n_4\;.
\earr \label{eq.prox22}
\ee
For example, $\{1\} \rightarrow \{2\}$, $\{2\} \rightarrow \{4\}, \{2,2\}$, $\{3\} \rightarrow \{6\}, \{4,2\}, \{2,2,2\}$,
$\{4\} \rightarrow \{8\}, \{6,2\}, \{4,4\}, \{4,2,2\}, \{2,2,2,2\}$, $\{5\} \rightarrow \{10\}, \{8,2\}, \{6,4\}, \{6,2,2\}, \{4,4,2\}$, $\{4,2,2,2\}$.
However, as discussed in Section 3.1, the $\{f\}$ of $U(10)$ are four columned and therefore we need to expand $\{f\}$ into Kronecker products of symmetric irreps of $U(10)$ and then apply Eq. (\ref{eq.prox22}). Expansion of $\{f\}$ of say $r$ number of rows into symmetric irreps is given by the determinant of the $r \times r$ matrix $g$ with,
\be
\{f\} = \l|g_{ij}\r|\;,\;\;\;g_{ij}=f_i+j-i,\;\;\{0\}=1,\;\;\{-x\}=0\;.
\label{eq.prox23}
\ee
Alternatively, it is also possible to expand $\{f\}$ in terms of totally antisymmetric irreps \cite{Wy-70,Little}. Using Eq. (\ref{eq.prox23}) we have for example,
$\{1^2\} = \{1\} \times \{1\} - \{2\}$, $\{44\}=\{4\} \times \{4\} - \{5\} \times \{3\}$, $\{42\} =\{4\} \times \{2\} - \{5\} \times \{1\}$. 
Similarly, for example $\{444\}$ involves $3 \times 3$ determinant, $\{4442\}$ involves $4 \times 4$ determinant and so on. Using the expansion in terms of symmetric $U(10)$ irreps and then applying Eq. (\ref{eq.prox22}) for each symmetric irrep of $U(10)$ will give sum of products of $U(4)$ irreps. 
Again each $U(4)$ irrep can be expanded into symmetric irreps of $U(4)$ using Eq. (\ref{eq.prox23}). Now, evaluation of the Kronecker products of $U(4)$ irreps follows from the rules given for example in \cite{Wy-70,Little} and simplified in \cite{Nova}. 
Using this for example it is easy to obtain
$\{1^2\}_{U(10)} \rightarrow \{31\}_{U(4)}$.
Once we have the $U(4)$ irreps, they can be converted into $SO(6)$ irreps using Eq. (\ref{eq.prox11a}). For example,
$\{4\}_{U(4)} \rightarrow [2,2,2]_{SO(6)}$,
$\{22\}_{U(4)} \rightarrow [2]_{SO(6)}$ and
$\{31\}_{U(4)} \rightarrow [2,1,1]_{SO(6)}$.

An alternative that gives easily the leading $U(4)$ or $SO(6)$ irrep, is to use the result that $U(4)$ corresponds to 4-dimensional oscillator and its basic irrep $\{2\}$ correspond to oscillator shell with two quanta. Then, the sp orbits
are 10 in number and in Cartesian co-ordinates they are
$\{n^i_1,n^i_2,n^i_3,n^i_4\}$. In the order $i=1,2,\ldots,10$ these are
$$
(2000)_1, (1100)_2, (1010)_3, (1001)_4, (0200)_5, (0110)_6, (0101)_7, (0020)_8, (0011)_9, (0002)_{10}
$$
Now, given a $U(10)$ irrep $\{f\}=\{f_1,f_2,\ldots,f_{10}\}$, the leading or the highest weight (h.w.) $U(4)$ irrep is obtained by putting $f_i$ number of particles in the $\{n_1,n_2,n_3,n_4\}_i$ orbit giving
\be
\l\{f_1,f_2,\ldots,f_{10}\r\} \rightarrow \{\cf\}_{h.w.} = \l\{\sum_i f_i n^i_1, \sum_i f_i n^i_2, \sum_i f_i n^i_3, \sum_i f_i n^i_4\r\}\;.
\label{eq.prox24}    
\ee 
Now, Eq. (\ref{eq.prox11a}) gives the h.w. $SO(6)$ irrep $\l[\sigma_1,\sigma_2,\sigma_3\r]_{h.w.}$.
For example we have,
$$
\barr{l}
\{4,2\}_{U(10)} \rightarrow \{10,2\}_{h.w.:U(4)} \leftrightarrow
\l[6,4,4\r]_{h.w.:SO(6)}\;,\\
\{4,4\}_{U(10)} \rightarrow \{12,4\}_{h.w.:U(4)} \leftrightarrow
\l[8,4,4\r]_{h.w.:SO(6)}\;,\\
\{4,4,4\}_{U(10)} \rightarrow \{16,4,4\}_{h.w.:U(4)} \leftrightarrow \l[8,8,4\r]_{h.w.:SO(6)}\;,\\
\{4,4,4,2\}_{U(10)} \rightarrow \{18,4,4,2\}_{h.w.:U(4)} \leftrightarrow \l[8,8,6\r]_{h.w.:SO(6)}\;.
\earr
$$
Reduction of a $SO(6)$ irrep $\l[\sigma_1,\sigma_2,\sigma_3\r]$ to $SO(5)$ irreps $\l[v_1,v_2\r]$ follows from the simple rule,
\be
\l[\sigma_1,\sigma_2,\sigma_3\r]_{SO(6)} \rightarrow \dis\sum \l[v_1,v_2\r]_{SO(5)} \oplus \;;\;\;\;\;
\sigma_1 \ge v_1 \ge \sigma_2 \ge v_2 \ge \l|\sigma_3\r|\;.
\label{eq.prox25}
\ee
Then, for example
\be
\barr{l}
\l[2\r]_{SO(6)} \rightarrow \{\l[0\r], \l[1\r], \l[
2\r]\}_{SO(5)}\;,\;\;\;
\l[2,2,2\r]_{SO(6)} \rightarrow \l[2,2\r]_{SO(5)}\;,\\
\l[2,1,1\r]_{SO(6)} \rightarrow \{\l[2,1\r], \l[1,1\r]\}_{SO(5)}\;,\;\;
\l[6,4,4\r]_{SO(6)} \rightarrow \{\l[6,4\r], \l[5,4\r], \l[4,4\r]\}_{SO(5)}\;,\\
\l[8,8,4\r]_{SO(6)} \rightarrow \{\l[8,8\r], \l[8,7\r], \l[8,6
\r], \l[8,5\r], \l[8,4\r]\}_{SO(5)}\;.
\earr \label{eq.prox6t5}
\ee
Finally, the reduction of $SO(5)$ irreps $\l[v_1,v_2\r]_{SO(5)}$ into $(L)_{SO(3)}$ follow from the rules given in \cite{Wy-70,Nova} and some examples from the tables in \cite{Wy-70} are,
\be
\barr{l}
\l[0\r] \rightarrow (0);\;\;\l[1\r] \rightarrow (2);\;\;\l[2\r] \rightarrow (2),(4);\;\;\l[11\r] \rightarrow (1),(3); \\
\l[21\r] \rightarrow (1), (2), (3), (4), (5);\;\;\l[22\r] \rightarrow (0), (2), (3), (4), (6)\;;\\
\l[44\r] \rightarrow (0), (2), (3), (4)^2, (5), (6)^2, (7), (8), (9), (10), (12) \;;\\
\l[54\r] \rightarrow (1), (2)^2, (3)^2, (4)^3, (5)^3, (6)^3, (7)^3, (8)^3, (9)^2, (10)^2, (11)^2, (12), (13), (14) \;;\\
\l[64\r] \rightarrow (0), (1), (2)^3, (3)^2, (4)^4, (5)^4, (6)^5, (7)^4, (8)^5, (9)^4, (10)^4, \\
(11)^3, (12)^3, (13)^2, (14)^2, (15), (16)\;;\\
\l[84\r] \rightarrow (0), (1)^2, (2)^4, (3)^4, (4)^6, (5)^6, (6)^7, (7)^7, (8)^8, (9)^7, (10)^8, \ldots \;;\\
\l[86\r] \rightarrow (0), (1), (2)^3, (3)^3, (4)^5, (5)^5, (6)^6, (7)^6, (8)^7, (9)^6, (10)^7, \ldots\;.
\earr \label{eq.prox5t3}
\ee 
Note that $(L)^r$ above means the $L$ appears $r$ times (multiplicity of the irrep $L$). It is important to note that in general multiplicities appear in irrep reductions.

\subsection{Proxy-$SU(5)$ limit: irrep reductions}

Given a symmetric irrep $\{n\}$ of $U(10)$, the irreps $\{\ca\}$ of $U(5)$ follow easily from \cite{Little} as $\{1\}_{U(10)} \rightarrow \{1^2\}_{U(5)}$. Firstly, given an integer $n$ generate all the partitions $\{n_1,n_2,\ldots\}$ with $n=\sum_i n_i$ and $n_i \ge n_{i+1}$ with all $n_i \ge 0$. Now, $\{\ca\}$ are transpose (rows changed to columns) of the partitions $\{2n_1,2n_2,\ldots\}$. Then,
\be
\{n\}_{U(10} \rightarrow \{\ca\}_{U(5)} = \dis\sum_{\{n_1, n_2, \ldots \}} \;Trans{\l\{2n_1, 2n_2, \ldots \r\}} \oplus \;.
\label{eq.prox26}
\ee
Here, $Trans\{---\}$ denotes transpose of the irrep $\{---\}$. 
As a $U(5)$ irrep will have maximum of 5 rows, it is easy to see
that the irreps $\{2n_1, 2n_2, \ldots \}$ must be of the type
$\{4^a 2^b\}$ with $4a+2b=2n$. Therefore,
\be
\{n\}_{U(10)} \rightarrow \{\ca\}_{U(5)} = \dis\sum_{2a+b=n} Trans{\l\{4^a 2^b\r\}} = \dis\sum_{a \ge 0, a \le n/2} \l\{n-a,n-a,a,a\r\}\;.
\label{eq.prox26a}
\ee
Then for example,
\be
\barr{l}
\{2\}_{U(10)} \rightarrow \{\ca\}_{U(5)} = \l\{2,2\r\}, \l\{1,1,1,1\r\} \;,\\
\{3\}_{U(10)} \rightarrow \{\ca\}_{U(5)} = \l\{3,3\r\}, \l\{2,2,1,1\r\} \;,\\
\{4\}_{U(10)} \rightarrow \{\ca\}_{U(5)} = \l\{4,4\r\}, \l\{3,3,1,1\r\}, \l\{2,2,2,2\r\} \;,\\
\{5\}_{U(10)} \rightarrow \{\ca\}_{U(5)} = \l\{5,5\r\}, \l\{4,4,1,1\r\}, \l\{3,3,2,2\r\} \;,\\
\{n\}_{U(10)} \rightarrow \{\ca\}_{U(5)} = \l\{n,n\r\}, \l\{n-1,n-1,1,1\r\}, \l\{n-2,n-2,2,2\r\}, \ldots \;.
\earr \label{eq.prox26b}
\ee
Now, just as in Section 5.1, any given $\{f\}$ of $U(10)$ can be reduced to $U(5)$ irreps by expanding $\{f\}$ in terms of Kronecker products of symmetric $U(10)$ irreps and then applying Eq. (\ref{eq.prox26}) gives sum of Kronecker products of $U(5)$ irreps. Again expanding these in terms of symmetric irreps of $U(5)$ and carrying out the Kronecker products will give $\{f\} \rightarrow \{\ca\}$. For irreps of small integers, this is easy to use. For example using this gives $\{1^2\}_{U(10)} \rightarrow
\{211\}_{U(5)}$. In general this algebra is tedious (let us mention that a computer code for evaluating various Kronecker products is available \cite{Schur}) but it can be used to obtain the leading or h.w. $U(5)$ irrep in a given $\{f\}$ of $U(10)$. For example,
\be
\barr{l}
\{4,2\}_{U(10)} \rightarrow \{6,4,2\}_{h.w. :U(5)} \;,\\
\{4,4\}_{U(10)} \rightarrow \{8,4,4\}_{h.w. :U(5)} \;,\\
\{4,4,2\}_{U(10)} \rightarrow \{10,4,4,2\}_{h.w. :U(5)} \;.
\earr \label{eq.prox26c}
\ee
From these it is possible to guess that $\{4,4,4\}_{U(10)} \rightarrow \{12,4,4,4\}_{h.w. :U(5)}$ and similarly,
$\{4,4,4,2\}_{U(10)} \rightarrow \{14,4,4,4,2\}_{h.w. :U(5)}$. 
Going further, one can use the rules given in \cite{Wy-70,Little} and obtain reductions of $\{\ca\}$ to irreps $\l[v_1,v_2\r]$ of $SO(5)$; see Eq. (\ref{eq.prox27}) ahead. Also, the irrep reductions for $\{\ca\} \rightarrow \l[v_1,v_2\r]$ for many cases of interest are tabulated in \cite{Wy-70}. For example, using these tables we have,
\be
\barr{l}
\{6,4,2\}_{U(5)} \rightarrow \l[v_1,v_2\r]_{SO(5)} = \l[0\r], \l[2\r]^3, \l[22\r]^3, \l[31\r]^3, \l[32\r]^2, \l[33\r], \l[4\r]^3, \l[41\r], \l[42\r]^5, \l[43\r]^2, \\
\l[44\r]^2, \l[51\r]^2, \l[52\r]^2, \l[53\r]^2, \l[54\r], \l[6\r], \l[62\r]^2, \l[63\r], \l[64\r]\;.\\
\{8,4,4\}_{U(5)} \rightarrow \l[v_1,v_2\r]_{SO(5)} = \l[0\r], \l[2\r]^2, \l[22\r]^2, \l[31\r], \l[32\r], \l[4\r]^3, \l[42\r]^4, \l[43\r], \l[44\r]^3, \l[51\r]^2, \\ 
\l[52\r]^2, \l[53\r]^2, \l[54\r]^2, \l[6\r]^2, \l[62\r]^3, \l[63\r], \l[64\r]^2, \l[71\r], \l[72\r], \l[73\r], \l[74\r], \l[8\r], \l[82\r], \l[84\r]\;.
\earr \label{eq.prox5t5}
\ee
Further, $\l[v_1,v_2\r] \rightarrow (L)$ reductions are tabulated in \cite{Wy-70} for all cases with $v_1 \le 8$. These are used in
Eq. (\ref{eq.prox5t3}) given before.  
 
\subsection{Proxy-$SO(10)$ limit: irrep reductions}

Reduction of the irreps $\{f\}$ of $U(10)$ to $SO(10)$ irreps $\l[\omega\r]$ follows from the general rule given in \cite{Wy-70,Little} and used in \cite{KoCas}. This rule is,
\be
\barr{l}
\{f\}_{U(10)} = \dis\sum_{\{\omega\}} \dis\sum_{\{\delta\}} \Gamma_{\{\delta\} \,\{\omega\}\,\{f\}}\,\l[\mu\r]_{SO(10)}\;,\\
\{\delta\} = \{0\}, \{2\}, \{4\}, \{2^2\}, \{6\}, \{42\}, \{2^3\}, \{8\}, \ldots\;.
\earr \label{eq.prox27}
\ee
In Eq. (\ref{eq.prox27}), $\Gamma_{\{\delta\} \,\{\omega\}\,\{f\}}$ is the multiplicity of $\{f\}$ in the Kronecker product $\{\delta\} \times \{\mu\} \rightarrow \{f\}$. Note that the sums above are direct sums and also $\{\delta\}$ involve only even entries.
There are additional rules when $\l[\omega\r]$ has more than five rows \cite{Wy-70,Little}. Applying Eq. (\ref{eq.prox27}), we have for example
\be
\barr{l}
\{2\} \rightarrow \l[0\r] + \l[2\r]\;,\;\;\;\;\{1^2\} \rightarrow \l[11\r]\;,\;\;\;\;\{4\} \rightarrow \l[0\r] + \l[2\r] + \l[4\r]\;,\\
\{42\} \rightarrow \l[0\r] + \l[2\r]^2 + \l[22\r] + \l[31\r] + \l[4\r] + \l[42\r] \;,\\
\{44\} \rightarrow \l[0\r] + \l[2\r] + \l[22\r] + \l[4\r] + \l[42\r] + \l[44\r]\;,\\
\{442\} \rightarrow \l[0\r] + \l[2\r]^2 + \l[22\r]^2 + \l[222\r] + \l[31\r] + \l[321\r] + \l[4\r] + \l[42\r]^2 + \l[422\r] + \ldots \;, \\
\{444\} \rightarrow \l[0\r] + \l[2\r] + \l[22\r] + \l[222\r] + \l[4\r] + \l[42\r] + \l[422\r] + \l[44\r] + \l[442\r] + \l[444\r]\;,\\
\{4442\} \rightarrow \l[0\r] + \l[2\r]^2 + \l[22\r]^2 + \l[222\r]^2 + \l[2222\r] + \l[31\r] + \l[321\r] + \l[3221\r] + \l[4\r] + \l[42\r]^2 + \ldots\;.
\earr \label{eq.prox28}
\ee
Finally, reductions of $\l[\omega\r]$ of $SO(10)$ to the $SO(5)$ irreps $\l[v_1,v_2\r]$ follow by starting with the $\{f\}$ for 1, 2, 3, $\ldots$ particles and then obtain $\{f\} \rightarrow \l[v_1,v_2\r]$ using the procedure outlined in Section 5.1. Using this and $\{f\} \rightarrow \l[\omega\r]$ given by Eq. (\ref{eq.prox27}), it is possible to obtain by a substation procedure $\l[\omega\r] \rightarrow \l[v_1,v_2\r]$. Let us add that $\l[v_1,v_2\r] \rightarrow L$ for $v_2=0$, 1 and 2 follow easily from the procedure given for example in \cite{Nova}. Similarly, as mentioned before reductions for all $\l[v_1,v_2\r]$ with $v_1 \le 8$ are tabulated in \cite{Wy-70}. 

\subsection{Ground state structures}

In order to apply the four symmetry limits $SU^{\alpha}(3)$,
$SO^{\alpha}(6)$, $SU^{\alpha}(5)$ and $SO^{\alpha}(10)$ with good proxy-$SU(4)$ symmetry, as a first step, it is important to understand the ground state structures and the typical spectra generated by these limits. These are briefly discussed here using the results in Sections 4 and 5.1-5.3. 

Firstly, it is easy to see that $SU^{(-)}(3)$ limit generates prolate shape and $SU^{(+)}(3)$ limit generates oblate shape. As discussed in detail in \cite{Kota-1}, this follows from the
shell model adopted quadrupole transition operator $T^{E2}$ and it is same as $Q^2_q(\alpha)$ operator in Eq. (\ref{eq.proxc}) with $\alpha=-1$. Going further, it is easy to infer from the previous studies on pairing
with LST coupling, that the quadratic Casimir operator of $SO^{\alpha}(10)$ is essentially same as isoscalar ($L=0,S=1,T
=0$) plus isovector ($L=0,S=0,T=1$) pairing Hamiltonian with both parts having equal strength \cite{Pang,KoCas}. Note that each of these pair operators are sum of pair operators in $\ell=1$ and with a factor $\beta(=\pm 1)$, in $\ell=3$ spaces. Also, $\alpha$ is related in a simple manner to the phase $\beta$ in the pairing operators \cite{KoCas}.

Ground state structure and the typical spectrum generated by the
$SO^{\alpha}(6)$ and $SU^{\alpha}(5)$ symmetry limits can be
obtained for example using the Hamiltonian (a linear combination of the quadratic Casimir operators of the algebras involved), 
\be
\barr{rcl}
H & = & \alpha_1\,\l[-\cc_2(U(10))\r] + \alpha_2\,\l[-\cc2(SO^{\alpha}(6))\r] + \alpha_3\,\l[-\cc_2(SO^{\alpha}(5))\r] \\
& + & \alpha_4\,\l[-\cc_2(SU^{\alpha}(5))\r] + \alpha_5\,\l[-\cc_2(SO(3))\r] \;.
\earr \label{eq.h1}
\ee
Note that the first term with $\alpha_1 > 0$ assures that the low-lying states of a N=Z even-even nucleus with number of nucleons $m=4r$ belong to the $U(10)$ irrep $\{4^r\}$ (this gives $(S,T)=(0,0)$) and that of a N=Z odd-odd nucleus with $m=4r+2$ belong to the irrep $\{4^r2\}$ (this gives $(ST)=(10)$ and $(01)$). Similarly, the last term with $\alpha_5 < 0$ gives 
the state with $L=0$ lowest. More importantly, the $H$ with $\alpha_4=0$ will generate states with $SO^{\alpha}(6)$ symmetry.
In this situation, $\alpha_2 >0$ gives states with highest $SO(6)$ irrep lowest and similarly $\alpha_3 <0$ gives states with lowest $SO(5)$ irrep lowest. On the other hand, the $H$ with $\alpha_2=0$ gives states with $SU^{\alpha}(5)$ symmetry and here clearly we need to use $\alpha_3 < 0$ and $\alpha_4 > 0$. With these, for example, for a system with 4 valence protons and 4 valence neutrons, low-lying levels in the $SO(6)$ limit are labeled by the $U(10)$ irrep $\{4^2\}$ (with $(S,T)=(0,0)$), $SO(6)$ irrep $\l[844\r]$ and $SO(5)$ irreps $\l[44\r]$, $\l[54\r]$, $\l[64\r]$, $\ldots$ in that order (see Section 5.1). The $L$ values in the $SO(5)$ irreps [see Eq. (\ref{eq.prox5t3})] can be arranged into some band structures. Thus, the ground state in this example is
$$
\l.\l| \{44\}_{U(10)}, \l[844\r]_{SO(6)}, \l[44\r]_{SO(5)}, L=0,S=0,T=0\r.\ran\;.
$$
As already mentioned in Section 4, spectra in the symmetry limits do not depend on $\alpha$. However, the wavefunction structure depends on $\alpha$ \cite{Kota-1,Kota-2}. With $SO(5)$ present between $SO(6)$ and $SO(3)$, clearly there will be shape mixing and possibly shape coexistence in this symmetry limit. This is also brought out by a preliminary study we have carried out using DSM based on HF sp states with axial symmetry. These show [obtained using Eq. (\ref{eq.h1}) with $\alpha_4=0$] that the ground state is a mixture of prolate and oblate shapes and the mixing depends on $\alpha$. Turning to the $SU(5)$ limit, for the 4 proton and 4 neutron example, the ground state irreps follow from the results in Section 5.2 giving,
$$
\l.\l| \{44\}_{U(10)}, \{844\}_{SU(5)}, \l[0\r]_{SO(5)}, L=0,S=0,T=0\r.\ran\;.
$$
Note that the irrep $\{844\}$ of $SU(5)$ gives large number of $SO(5)$ irreps (see Eq. (\ref{eq.prox5t5})) and the various $L$ states that belong to each of these again can be arranged into some band structures. Clearly, the structure here will be different from the $SO(6)$ limit. DSM calculations using $H$ in Eq. (\ref{eq.h1}) with $\alpha_2=0$ also showed that the wavefunction structure in the $SU(5)$ limit is quite different from $SO(6)$ limit. More detailed studies using not only DSM but also the shell model codes will be reported elsewhere.

\section{Conclusions and future outlook}

Proxy-$SU(3)$ model for nuclei in A=60-90 region introduces automatically proxy-$SU(4)$ symmetry. Shell model spaces with sp orbits $^1p_{3/2}$, $^1p_{1/2}$, $^0f_{5/2}$ and $^0g_{9/2}$ are essential for these nuclei and also protons and neutrons in this region occupy the same sp orbits. Therefore, with the proxy scheme, the SGA is $U(40) \otimes SU_{ST}(4)$. With the spatial $U(10)$ algebra containing $SU(3)$ subalgebra, we have the proxy-$SU(3)$ model of Bonatsos et al. Shell model calculations pointing out the need for $^0g_{9/2}$ orbit, ground state masses, shape changes and shape co-existence in A=60-90 region clearly show the importance of proxy-$SU(4)$ in this mass region. These are described in Sections 2 and 3. Besides admitting $SU(3)$ subalgebra, the $U(10)$ algebra admits three new proxy subalgebras with $G=SU(5)$, $SO(6)$ and $SO(10)$ all with good proxy-$SU(4)$ symmetry. In addition, these subalgebras appear twice. For these three subalgebras we have presented in Sections 4 and 5, the group generators, quadratic Casimir invariants and their eigenvalue formulas and also results for various group irrep reductions. Also, ground state structure generated by them is discussed briefly.

Going further, it is necessary to address the following in future. (i) Develop Wigner-Racah algebra for the three new $SO(6)$, $SU(5)$ and $SO(10)$ algebras. This will allow us to derive results for various transition matrix elements in the symmetry limits. However this appear to be an uphill task. (ii) An alternative is to perform large scale shell model calculations using various combinations of the Casimir invariants of the various groups in $U(10)$ subalgebras and also using general multipole-multipole interactions ($Q^2 \cdot Q^2$, $Q^3 \cdot Q^3$ etc.) constructed in $U(10)$ space. (iii) Using (i) or (ii), obtain predictions for electro-magnetic transition strengths, GT strengths, number of $pp$, $nn$ and $pn$ pairs in the ground states and so on generated by the new $SO(6)$, $SU(5)$ and $SO(10)$ symmetry limits and also with some mixing of these algebras with $SU(3)$. These are needed for experimental data analysis. (iv) Understand the differences between the two $SU(3)$, $SO(6)$, $SU(5)$ and $SO(10)$ algebras; see \cite{Kota-1,Kota-2} for previous studies on multiple pairing and $SU(3)$ algebras in IBM and shell model. (v) Understand the role of three and higher order Casimir operators of $SU(3)$, $SO(6)$, $SU(5)$ and $SO(10)$, $SO(5)$ and $SU_{ST}(4)$ in A=60-90 region; see \cite{Casten} for showing the importance of 3-body forces in A=60-90 region. 

Finally, let us add that just as the $SO(6)$ (or $SU(4)$) and $SU(5)$ algebras in addition to $SU(3)$ in $U(10)$ of $\eta=3$ shell, it is possible to consider in $\eta=4$ shell, $SU(5)$ (with symmetric irrep $\{2\}$ giving $\ell=0,2,4$) and $SU(6)$ (with $\{1^2\}$ irrep giving $\ell=0,2,4$) symmetry limits in $U(15)$ besides $SU(3)$ for identical nucleons. Similarly, there will be $SU(6)$ and $SU(7)$ algebras in $U(21)$ of $\eta=5$ shell and $SU(7)$ and $SU(8)$ in $U(28)$ of $\eta=6$ shell. All these symmetry limits will extend the scope of proxy-$SU(3)$ model as 
applied to heavy nuclei. 

\ack

Thanks are due to P.C. Srivastava for useful correspondence. RS is thankful to SERB of DST (Government of India) for financial support.


\end{document}